\documentclass[aps,prl,reprint,showpacs]{revtex4-2}

\usepackage{graphicx}
\usepackage{dcolumn}
\usepackage{bm}

\usepackage{cases}
\usepackage{eqnarray}

\usepackage{amssymb}

\usepackage{slashed}

\usepackage{color}

\usepackage{subfigure}
\usepackage{multirow}
\usepackage{natbib}

\graphicspath{{figs/}{figsgaoerb/}} 

\begin{document}

\title{ Existence of an optimized stellarator with simple coils}

\author{Guodong Yu}
\author{Zhichen Feng}
\author{Peiyou Jiang}
\author{GuoYong Fu}\thanks{corresponding author's Email: gyfu@zju.edu.cn}
\affiliation{Institute for Fusion Theory and Simulation and Department of physics, Zhejiang University, Hangzhou 310027, China}

\begin{abstract}
An optimized compact stellarator with four simple coils is obtained from direct optimization via coil shape. The new stellarator consists of two interlocking coils 
and two vertical field coils similar to those of the Columbia Non-neutral Torus (CNT)[Pedersen et al. Phys. Rev. Lett. 88, 205002 (2002)]. The optimized 
configuration has global magnetic well and a low helical ripple level comparable to that of Wendelstein 7-X (W7-X)[Wolf et al. Nucl. Fusion 57, 102020 (2017)]. 
The two interlocking coils have a smooth three-dimensional shape much simpler than those of advanced stellarators such as W7-X. This result opens up possibilities of future stellarator reactors with simplified coils. 
\end{abstract}

\maketitle

The two main approaches of magnetic fusion energy (MCF) are tokamak's and stellarator's. Tokamak is currently the dominant 
approach with advantages of axisymmetric geometry and achieved plasma parameters significantly better than those of other 
MCF devices. However, stellarators have recently enjoyed a renaissance as recent results of the advanced stellarator Wendelstein 7-X (W7-X)\cite{Wolf2017} demonstrated the reduced neoclassical energy transport\cite{Dinklage2018}\cite{Beidler2021}. 
It is expected that plasma performance of W7-X can reach a level comparable to that of an equivalent tokamak in next few years. 
Stellarators have advantages of naturally steady state operations because their magnetic fields are mainly generated by external coils and plasma current is not needed for confinement and is usually quite small. The harmful current-driven instabilities such as disruptions in tokamaks are absent in stellarators. The confinement properties of stellarator plasmas are largely determined by external coils. The recent success of W7-X demonstrates that it is possible to build the optimized three dimensional (3D) coils to such high precisions that the designed plasma confinement performance can be achieved as predicted in actual experiments. It has been argued recently that the stellarator approach provides the fastest track to the realization of fusion energy because favorable plasma confinement properties can be designed, realized and controlled almost fully by external 3D coils\cite{Boozer2021}. 

The advanced stellarators such as W7-X are usually designed for improved neoclassical confinement and MHD stability by two stage approaches. First, the shape of plasma boundary is varied to obtain optimized properties. Second, the optimized boundary shape is realized by designing the shapes of 3D coils. Unfortunately, the resulting 3D coils are usually quite complex and are difficult and costly to build. This is evidenced by the long delay of W7-X project \cite{Ribe2009} and the cancellation of the NCSX project \cite{NEILSON2010} due to difficulties in engineering and building of 3D coils. Therefore it is necessary to simplify the 3D coils for stellarator to become an economical platform for fusion reactors. For this reason, recently, coil simplification of stellarator has been a subject of intensive studies including reducing coil complexity in the optimization\cite{LOBSIEN2018} \cite{Kruger2021}\cite{Henneberg2021} as well as using permanent magnets to replace 3D coils\cite{HELANDER2020}\cite{Zhu2020}\cite{Xu2021}. 

In this work, we explore the possibilities of optimized stellarators with simplified coils. We use a direct optimization method of varying coil shape to optimize both plasma confinement and MHD stability. In this way we can effectively control the complexity of coils while optimizing the plasma properties. The details of optimization method will be given later.  Following our recent 
work\cite{Yu2021}, we choose the coil topology of the Columbia Non-neutral Torus (CNT)\cite{Pedersen2012} for our optimization space. 
CNT consists of two circular interlocking(IL) coils and two circular vertical field (VF) coils. It is arguably the simplest compact stellarator ever built and successfully operated\cite{Pedersen2012}. Here we carry out a global optimization of both neoclassical confinement and MHD stability by varying the 3D shape of the two IL coils. An initial phase of this work has been done recently 
by targeting only the neoclassical transport. The results showed that the effective helical ripple level was reduced by an order of magnitude as compared to that of CNT indicating significant improvement in neoclassical confinement\cite{Yu2021}. This was realized by two planar IL coils with elliptical shape. In present work, a much broader parameter space of coil shape 
is explored allowing 3D shape of the two IL coils. As a result, an optimized Compact Stellarator with Simple Coils(CSSC), to be called CSSC, has been discovered. The CSSC has both global magnetic well and a low level of helical ripple comparable to that of W7-X. This breakthrough opens up possibilities of stellarator reactors with simplified coils. The design of CSSC can 
be used as basis for a low-cost experiment to study the physics of compact stellarators. The design could also be used as a candidate stellarator for electron-positron plasma experiments\cite{Pedersen2012}.

We now describe the configuration specification and physics properties of CSSC in details. 
In our optimization, the radius of the two circular VF coils is varied while the distance between them is fixed. The shapes of the two IL coils are varied and are described by following Fourier representation\cite{Zhu2017}

\begin{numcases}{}
x=x_{c,0}+\sum_{n=1}^{n_f}x_{c,n}\cos(nt),\\
y=\sum_{n=1}^{n_f}y_{s,n}\sin(nt),\\
z=\sum_{n=1}^{n_f}z_{s,n}\sin(nt),
\end{numcases}
where $(x,y,z)$ are Cartesian coordinates of line coils, $t$ is an angle variable ranges $[0,2\pi]$ and $n$ is the harmonic number.  The total number of Fourier coefficients for each IL coil
is $3\times n_f +1$ with $n_f$ being the maximum harmonic number. In order to maintain two periods of flux surfaces,  the two IL coils are constrained to have same shape. Furthermore, $n_f=3$ is chosen in this work. This choice of $n_f=3$ is a compromise between allowing enough degree 
of freedom while constraining the coil complexity. Finally, taking into account the current ratio between IL coils and VF coils, the optimization parameter space has a total of 12 degree of freedom. 

Table \ref{table1} shows the Fourier coefficients of the optimized configuration CSSC (4th column) along with the optimization ranges for each Fourier coefficient (3rd column).

\begin{table}[b]
\centering
\begin{ruledtabular}
\begin{tabular}{ccccc lll}

                                                                                & Parameter          & Range            & Results       &  &  &  \\ \colrule
\multirow{10}{*}{\begin{tabular}[c]{@{}c@{}}IL coil\end{tabular}}               & $x_{c,0}$          & 0.3$\sim$0.6     & 0.509         &  &  &  \\ 
                                                                                & $x_{c,1}$          & 0.6$\sim$1.2     & 1.051         &  &  &  \\ 
                                                                                & $x_{c,2}$          & -0.3$\sim$0.3    & 0.227         &  &  &  \\ 
                                                                                & $x_{c,3}$          & -0.1$\sim$0.1    & -2.622e-2     &  &  &  \\ 
                                                                                & $y_{s,1}$          & 0.6$\sim$1.2     & 0.806         &  &  &  \\ 
                                                                                & $y_{s,2}$          & -0.3$\sim$0.3    & 0.162         &  &  &  \\ 
                                                                                & $y_{s,3}$          & -0.1$\sim$0.1    & -6.257e-2     &  &  &  \\ 
                                                                                & $z_{s,1}$          & -0.4$\sim$0.8    & 0.667         &  &  &  \\ 
                                                                                & $z_{s,2}$          & -0.2$\sim$0.2    & -3.87e-2      &  &  &  \\
                                                                                & $z_{s,3}$          & -0.1$\sim$0.1    & -6.052e-2     &  &  &  \\ \colrule
\multirow{2}{*}{\begin{tabular}[c]{@{}c@{}}PF coil\end{tabular}}                & $y_{s,1}=x_{c,1}$  & 1.0$\sim$2.0     & 1.850         &  &  &  \\ 
                                                                                & $z_{c,0}$          & 0.77             & 0.770         &  &  &  \\ \colrule
Current ratio                                                                   & $I_{IL}/I_{PF}$    & 1$\sim$3         & 1.323         &  &  &  \\ 
\end{tabular}
\end{ruledtabular}
\caption{The optimization ranges of coil Fourier coefficients and coil current ratio are in the third column.
The optimized results are in the fourth column. }
\label{table1}
\end{table}

\begin{figure}
  \centering
  \includegraphics[height=4.5cm]{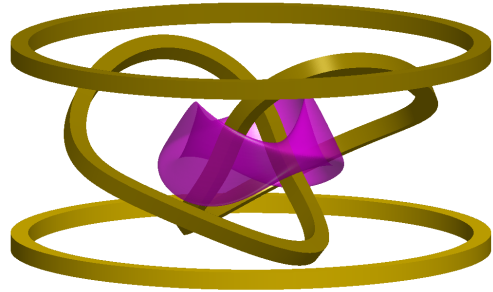}
  \caption{The optimized stellarator with four simple coils: the gold color denotes the two inner interlocked 3D  coils located between two circular vertical field coils; the purple color denotes the last closed flux surface.}\label{fig1}
\end{figure}

\begin{figure}[h]
  \centering
  
  \subfigure[$\phi=0^\circ$]{
  \begin{minipage}[t]{0.4\linewidth}
  \centering
  \includegraphics[width=1.1\textwidth,height=.24\textheight]{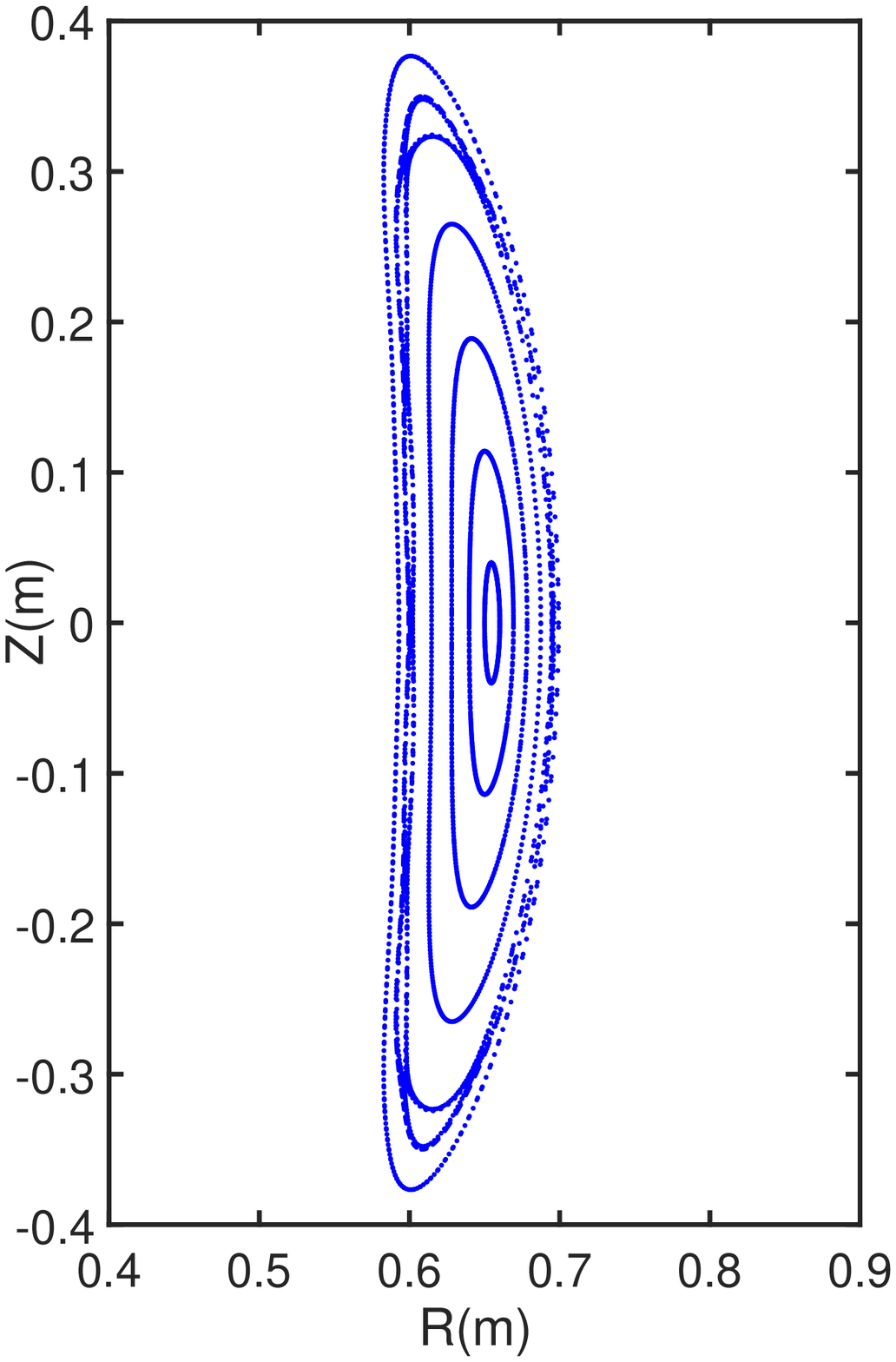}
  \end{minipage}
  }
  \subfigure[$\phi=90^\circ$]{
  \begin{minipage}[t]{0.4\linewidth}
  \centering
  \includegraphics[width=1.1\textwidth,height=.15\textheight]{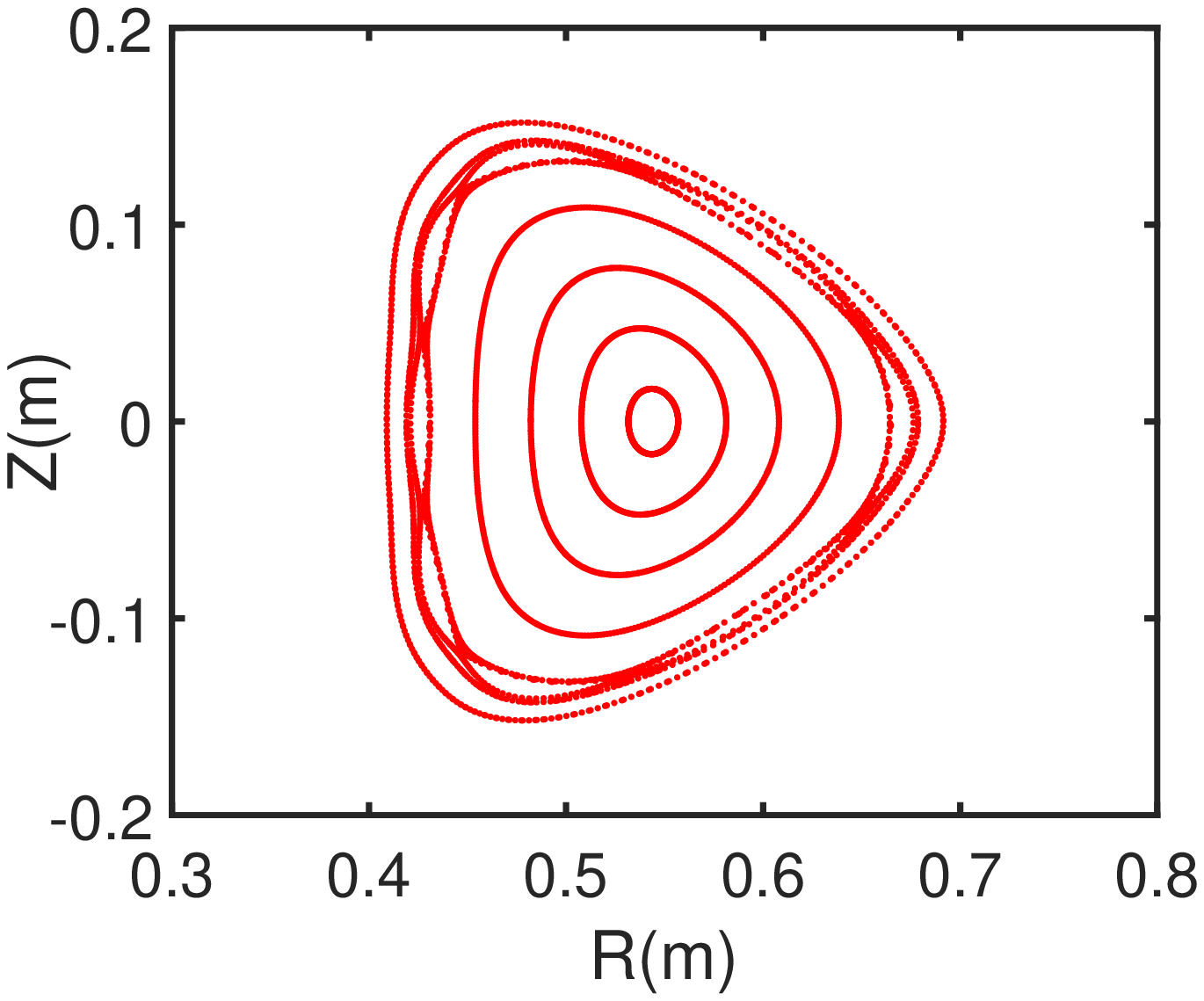}
  \end{minipage}
  }
  \caption{Poincare plot of vacuum magnetic surfaces at toroidal angels $0^\circ$ (left) and $90^\circ$ (right). 
    }\label{fig2}
\end{figure}

\begin{figure}
  \centering
  \includegraphics[height=6.5cm]{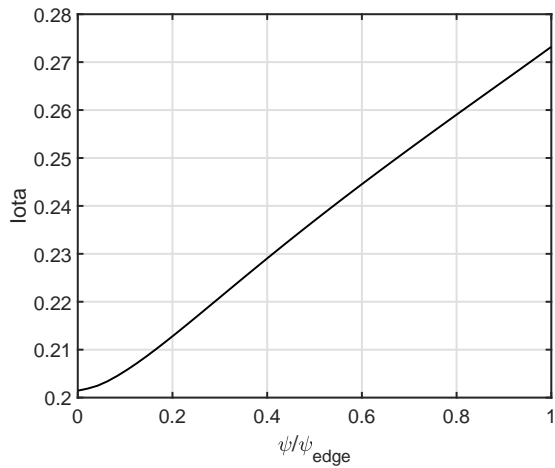}
  \caption{ Rotational transform iota versus normalized toroidal flux.
    }\label{fig3}
\end{figure}

Fig. \ref{fig1} shows the 3D schematic of coils and the last closed flux surface (LCFS) of CSSC. 
We observe that the two IL coils have a smooth shape with modest 3D variation. The averaged radius of the IL 
coils is $1.11m$. The radius of vertical field coils have a radius of $1.85m$. The distance between the two vertical 
field coils is $1.54m$. The current ratio of IL coils and VF coils is $1.323$. The minor and 
major radius of LCFS is $0.15m$ and $0.58m$ respectively, corresponding to a low aspect ratio of $R/a=3.87$. 
The volume of LCFS is $0.27m^3$. The Poincare plots of vacuum magnetic surfaces are shown in Fig. \ref{fig2} at toroidal 
angels $0^\circ$ and $90^\circ$ respectively. The configuration has good magnetic flux surfaces except for a small magnetic island at the $\iota=0.25$ surface. The rotational transform profile is shown in Fig. \ref{fig3} and is seen to vary from $0.2$ at the magnetic axis to $0.28$ at LCFS with modest magnetic shear. 

\begin{figure}
    \centering
    \includegraphics[height=6.5cm]{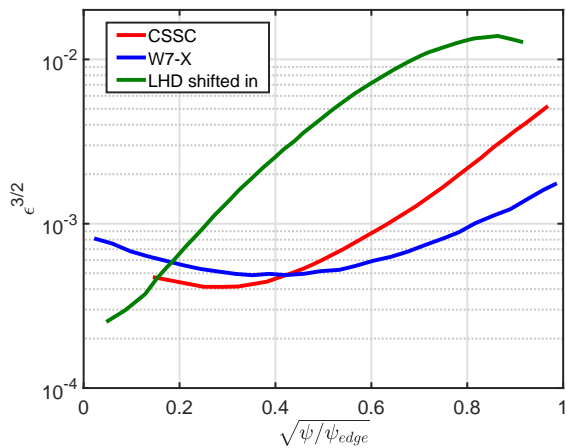}
    \caption{ Effective helical ripple $\epsilon_{eff}^{3/2}$ versus square root of the normalized toroidal flux $\sqrt{\psi/\psi_{edge}}$ for the optimized configuration (red line), W7-X (blue line)\cite{Beidler2021} and LHD (green line)\cite{Beidler2021}.
      }\label{fig4}
\end{figure}

One of our main optimization targets is neoclassical confinement, particularly the so-called $1/\nu$ neoclassical transport with $\nu$ being the collision frequency of plasmas. This $1/\nu$ scaling is very unfavorable for stellarator confinement for high temperature regime of fusion plasmas because the transport increases strongly with plasma temperature\cite{Darwin1987}. Therefore, the minimization of neoclassical transport in the $1/\nu$ regime is necessary for stellarator reactors. It has been shown that the $1/\nu$ transport 
is proportional the effective helical ripple coefficient $\epsilon_{eff}^{3/2}$\cite{Nemov1999}. We use this coefficient as a target. Fig. \ref{fig4} shows the radial profile of $\epsilon_{eff}^{3/2}$ of the 
CSSC as well as W7-X and the Large Helical Device (LHD)\cite{Iiyoshi1990}. We observe 
that $\epsilon_{eff}^{3/2}$ of CSSC is comparable to that of W7-X and is much smaller than that of LHD. 

\begin{figure}
  \centering
  \includegraphics[height=6.5cm]{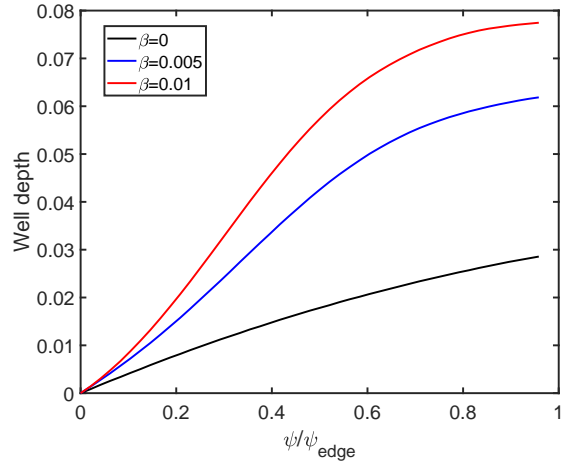}
  \caption{Magnetic well depth $W$ versus the normalized toroidal flux $\psi/\psi_{edge}$ for three values of plasma $\beta$. 
    }\label{fig5}
\end{figure}

The other important target of optimization is magnetic well which is a key parameter for MHD stability. The magnetic well is defined via the surface-averaged magnetic pressure\cite{Freidberg2014} as following:

\begin{equation}
  W=2\frac{V}{\left\langle B^2 \right\rangle } \frac{d}{dV} \left\langle \frac{B^2}{2} \right\rangle 
\end{equation}
where $B$ is magnetic field strength, $V(\psi)$ is the volume within flux surface $\psi$. A positive gradient of the surface-averaged magnetic field corresponds to a magnetic well. The magnetic 
well profile of CSSC is calculated with free boundary condition using the VMEC code for several plasma beta values. The results are plotted in Fig. \ref{fig5}. We see that CSSC possess a global magnetic well profile with well depth increasing with plasma beta. In addition to magnetic well, we have also carried out initial MHD stability calculation using the 3D MHD stability 
code TERPSICHORE \cite{Anderson1990} in order to confirm a more complete MHD stability of the CSSC. 
The results show that low-n global MHD modes are stable up to the volume-averaged beta of $\beta = 1\%$. It should be noted that our optimization is done for vacuum magnetic field. Future work will consider optimization at finite plasma beta.

\begin{figure}
  \centering
  \includegraphics[height=6.5cm]{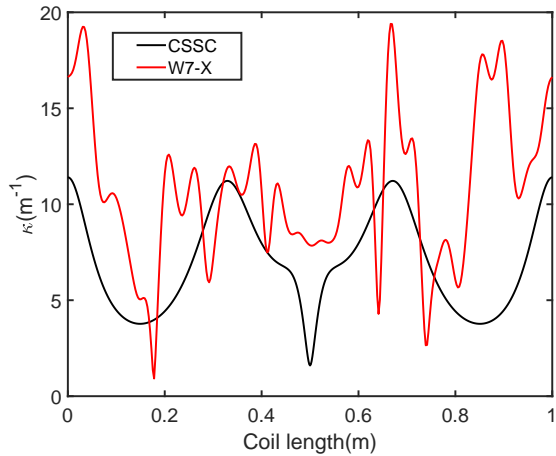}
  \caption{Coil curvature $\kappa$ variation along the loop of coil for the optimized configuration (black line) and W7-X (red).  
    }\label{fig6}
\end{figure}

So far, we have shown that the optimized configuration CSSC has both magnetic well and low level of helical ripple. Now we show that these favorable properties are achieved with relatively simple coils, i.e., the shape of the two IL coils is relatively smooth with modest 3D variation as compared to that of W7-X. Fig. \ref{fig6} compares the curvature of the optimized configuration with that of one of W7-X's module coils at the same coil length of $1m$. 
We observe that the coil curvature of CSSC is significantly smaller than that of the W7-X coil. Also 
the variation of coil curvature along the coil loop is much less complex indicating that the coil complexity of CSSC is much less in comparison.  

We now describe some details of the optimization method used to obtain above results. In addition to neoclassical transport and magnetic well, we also target plasma volume, rotational transform and flux surface quality. The magnetic field generated by current-carrying-coils is calculated from the Biot-Savart law. The flux surfaces and rotational transform profile are calculated by following the 
magnetic field lines. The effective ripple of each surface is calculated by integrating along a magnetic field 
line \cite{Nemov1999} as done in our recent work\cite{Yu2021}. 
We vary the shape of the two IL coils via the ten Fourier coefficients as described above to search for desirable configurations that meet our targets. Since the total degree of freedom is not a small number, a multi-stage random search algorithm is adopted in the optimization process.  Specifically, at the first stage, the initial ranges of each degree of freedom are selected as given 
in the TABLE \ref{table1}. It should be noted that the range of each Fourier coefficient is constrained to be smaller for higher $n$ in order to keep the coil complexity low. Given the ranges of parameters, a large collection of stellarator 
configurations are generated in the 12-dimension parameter space with value of each parameter randomly selected within its 
range. For the first stage, a total of $80$ million cases are evaluated. A few
good configurations are found that meet the criteria of $\epsilon_{eff}^{3/2}<0.01$, $\iota>0.2$, plasma volume $V>0.25m^3$, and having a global magnetic well profile. This concludes the first stage of search. 
At the second stage, this random search process is repeated starting from each of the good configurations found 
in the first stage. The ranges of parameters are updated based on each new configuration and are chosen to be narrower than those of the first stage. A total of four such stages of search have been carried out in arriving the final optimized 
configuration CSSC as defined in TABLE \ref{table1}. It should be noted that there are other optimized configurations with comparable quality. The details of other configurations will be given in another paper. 

In conclusion, a direct optimization from coils demonstrates the existence of a new optimized stellarator with four simple coils. The optimized configuration has favorable properties of magnetic well and a low effective helical ripple level comparable to that of W7-X. The two 3D interlocking coils have a smooth shape much simpler as compared to those of advanced stellarators such as W7-X.  This work opens up possibilities of future stellarator reactors with simplified coils. In near term, the optimized 
configuration can be used as a basis for a low-cost experiment to study the MHD stability and plasma confinement of compact stellarators. The design could also be used as a candidate stellarator for electron-positron plasma experiments.

\section*{Acknowledgement}

We thank Dr. Steve Hirshman for the use of the 3D equilibrium code VMEC code. We also thank Dr. W. A. Cooper for use of 
the 3D MHD stability code TERPSICHORE, and Dr. David Gates and Dr. Caoxiang Zhu for use of the stellarator optimization code STELLOPT. This work was funded by the start-up funding of Zhejiang University for one of the authors (Prof. Guoyong Fu).

\bibliographystyle{unsrt}

\end{document}